# Design of back-streaming white neutron beam line at CSNS

L.Y. Zhang[a,b,c], H.T. Jing[b,c*], J.Y. Tang[b,c,a], Q. Li[b,c], X. C. Ruan[d], J. Ren[d], C. J. Ning[b,c], Y. J. Yu[b,c], Z. X. Tan[b,c], P. C. Wang[b,c], Y. C. He[b,c], X.Q. Wang[a]

[a]*National Synchrotron Radiation Laboratory, University of Science and Technology of China, Anhui 230029, China*
[b]*Institute of High Energy Physics, Chinese Academy of Sciences (CAS), Beijing 100049,China*
[c]*Dongguan Neutron Science Center, Dongguan 523803,China*
[d]*China Institute of Atomic Energy, Beijing 102413, China*

**Abstract:** A white neutron beam line using back-streaming neutrons from the spallation target is under construction at China Spallation Neutron Source (CSNS). Different spectrometers, to be installed in the so-called Back-n beam line for nuclear data measurements, are also being developed in phases. The physical design of the beam line is carried out with the help of a complicated collimation system and a sophisticated neutron dump, taking the overview of the neutron beam characteristics into account. This includes energy spectrum, flux and time structure, the optimizations of neutron beam spots and in-hall background. The wide neutron energy range of 1 eV–100 MeV is excellent for supporting different applications, especially nuclear data measurements. At Endstation#2, which is about 80 m away from the target, the main properties of the beam line include neutron flux of $10^6$ n/cm$^2$/s, time resolution of a few per mille over nearly the entire energy range, and in-hall background of about 0.01 /cm$^2$/s for both neutron and gamma. With its first commission in late 2017, Back-n will not only be the first high-performance white neutron source in China, but also one of the best white neutron sources in the world.

## 1. Introduction

Complete nuclear data of good quality is very important for nuclear science and nuclear technology applications, such as designs and analyses of accelerator-driven subcritical systems, new-generation reactors, and nuclear medicine. At the same time, it is also required to solve some fundamental questions in nuclear physics and nuclear astrophysics. New generation white neutron sources, which are often associated with spallation neutron sources, can provide high neutron flux, very wide energy range and excellent energy resolution. Therefore, it opens new possibilities for high-precision cross section measurements. In recent years, several international neutron nuclear databases have gained rapid growth with the development of neutron sources, corresponding detectors and experimental technologies.

Apart from some single-energy accelerator-based neutron sources and thermal neutron sources from reactors, there are several white neutron sources in the world that are devoted to carry out neutron cross section measurements, e.g., n_TOF at the

European Organization for Nuclear Research (CERN), Weapons Neutron Research Facility (WNR) at Los Alamos Neutron Science Centre (LANSCE), Geel Electron Linear Accelerator (GELINA) at the Institute for Reference Materials and Measurements / Joint Research Centre (IRMM/JRC), and the Gaerttner LINAC Center at Rensselaer Polytechnic Institute (RPI) [1]. These facilities have specific characteristics of neutron beam intensities, energy resolutions, associated instrumentation and data acquisition systems, and dedicated experimental programs. The CERN n_TOF provides an intense pulsed neutron beam in a wide energy spectrum and a good energy resolution with a time-of-flight (TOF) distance of 185 m that stands today among the world's leading facilities for measuring neutron-induced cross section [2-3]. The LANSCE/WNR provides a continuous intense neutron beam with energy ranging from about 0.1 MeV to more than 600 MeV using the 800-MeV proton beam from the LANSCE linac and has six flight paths instrumented for a variety of measurements [4]. GELINA has an excellent neutron energy resolution for its short pulsed electron beam and long neutron flight path that ranges up to 400 m [5]. The RPI neutron source employs an electron Linear Accelerator (LINAC) with electron beam energy of 60 MeV to produce short neutron beam and its neutron measurement capability can cover an energy range from 0.005 eV to 20 MeV [6, 7].

China Spallation Neutron Source (CSNS), which is still under construction and expected to be completed in early 2018, is developed mainly for multidisciplinary research on material characterization by using neutron scattering techniques [8]. However, other kinds of applications at CSNS have also been planned and some of them have gained support from different resources. The Back-n project, which utilizes the back-streaming neutrons to build a white neutron source primarily for nuclear data measurements, has entered the construction since 2013 [9]. The white neutron beam is generated at the spallation target by a proton beam of 1.6 GeV and 100 kW. In this paper, the design of the Back-n beam line, including the layout and beam optics, energy spectrum, time resolution, and background suppression, is presented.

## 2. Characteristics of the back-streaming neutrons
### 2.1 Spallation target and proton beam line

A tantalum (0.3 mm) clad and water-cooled sliced tungsten target (11 pieces, 65 cm in total length, cross-section: $170 \times 70$ mm$^2$) is being used at CSNS. The target vessel is made of stainless steel (SS316) with a thickness of 2.5 mm for the front side, 7.5 mm for up and down sides and 12 mm for the lateral and back sides. The proton beam spot at the target entrance is of 12 cm (H) $\times 4$ cm (V) (four times the rms beam size) with a quasi-uniform distribution. The interaction between the pulsed high-energy proton beam and the heavy metal target produces a large number of neutrons through spallation reactions. Most of the neutrons are moderated by the target itself and three moderators, a Coupled Hydrogen Moderator, a Decoupled and Poisoned Hydrogen Moderator, and a Decoupled Water Moderator, (see Fig. 1) situating above and below the target. Thereafter, the moderated neutrons are transported to 19 neutron spectrometers for neutron scattering based application. However, there are also two white neutron beam lines in the proton beam plane: one is the Back-n beam line and the other is a 41° forward beam line (currently sealed and to be opened in the future) designated mainly for high-energy neutron irradiation studies. Between the target-moderator and the biological shielding structure, there are also reflectors of beryllium and iron to enhance the neutron utilization efficiency.

The Back-n project for nuclear data measurements is based on a previous report which describes the characteristics of the back-streaming neutrons [10]. The first 20-m beam line of the Back-n is shared with the last section of the proton beam line – RTBT (Ring to Target Beam Transport). A 2 mm thick aluminum alloy (A5083) proton beam window (PBW) is located at a distance of 1.9 m from the target separating the vacuum tube of the proton beam line and the target vessel. A copper collimator [length: 800 mm, aperture: 200 mm (H) × 100 mm (V)], which collimates the proton beam halo and also stops most of the back-streaming neutrons in large divergent angles, is placed at a distance of about 2.85 m from the target. Fig. 1 shows the geometry of the target area [11].

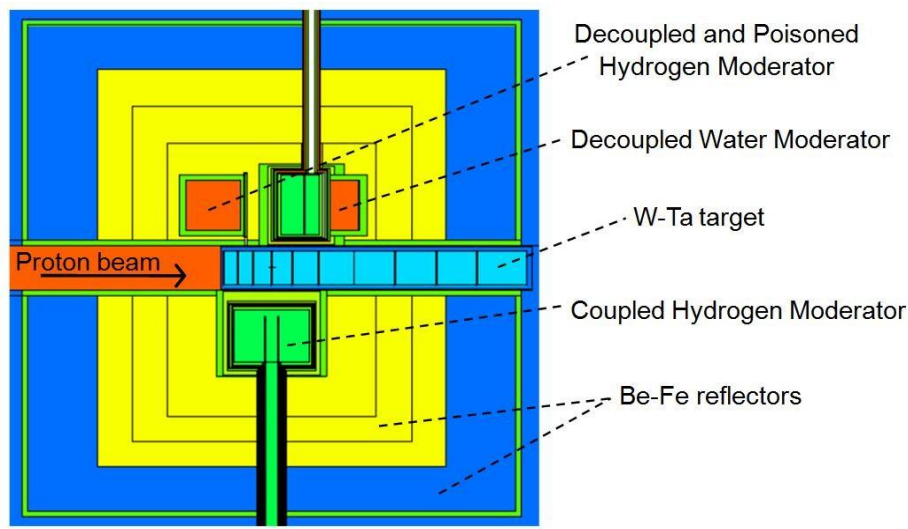

**Fig. 1 Geometry of target-moderator-reflector (TMR) model**

*2.2 Neutron energy spectrum*

The production, moderation, and transport of neutrons in the tungsten target have been simulated by using FLUKA code [12-13]. The simulation also takes the PBW into account. The energy spectra are slightly different at different locations in low-energy region because of the large scattering effect in target station. However, when the distance from the target surface is larger than 10 m, the energy spectra are similar. The simulated back-streaming neutron spectra at three important locations are shown in Fig. 2. The spectrum has a very wide energy-range, from the thermal neutron energy to several hundred MeV, and with a peak near 1 MeV, which is usually suitable for nuclear data measurements.

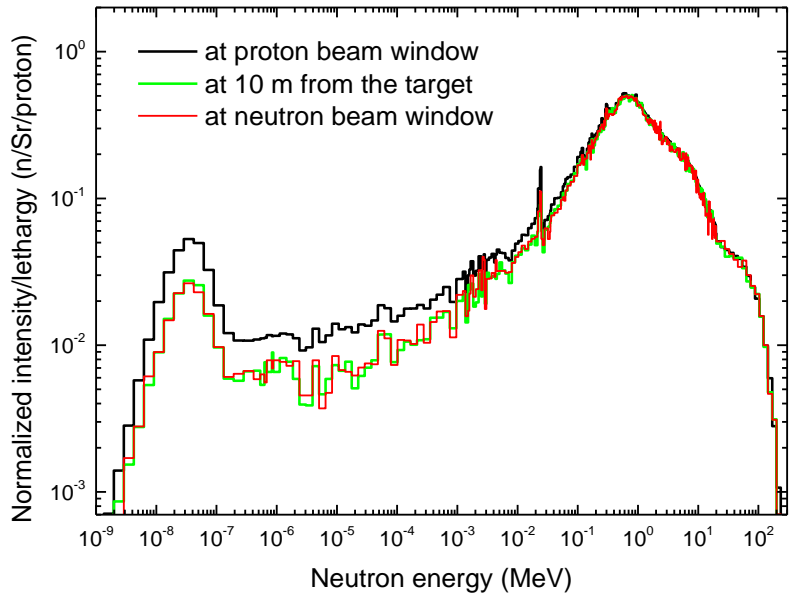

**Fig. 2 Back-streaming neutron energy spectrum from the CSNS target**

*2.3 Time resolution and CSNS operation modes*

Time resolution is a key parameter in the TOF method to determine neutron energy in a pulsed neutron source. Time resolution at Back-n will be affected by four main factors: 1) proton beam pulse structure defined by the accelerator; 2) neutron time expansion in the target where the neutrons are produced and moderated; 3) flight distance from the target to the neutron detector; 4) time error in neutron detection. The factors 2) and 3), mentioned above, are almost defined by the current CSNS target structure and Back-n layout. Factor 4) is strongly dependent on the type of the detector or spectrometer, and is not considered as part of the properties of a neutron beam line. Therefore, one can design special accelerator operation mode to shorten the proton pulse width and improve the time resolution of neutron measurement. We have developed different accelerator operation modes for this purpose, for example: 1) Normal Mode (N-mode): each proton pulse has two 13 ns-width (rms) bunches and are separated by 410 ns; 2) Dedicated Mode 1 (D-mode 1): only one bunch is kept in a proton pulse and the proton beam power is also halved; 3) Dedicated Mode 2 (D-mode 2): one bunch is used, but the rms bunch length is much reduced to only 1.5 ns, and the proton beam power is only 15% of the N-mode. The intrinsic neutron time resolution (factors 1)-3)) can be calculated by Monte Carlo simulations. On the other hand, it is found that for energy higher than 10 keV, the neutrons from two different proton bunches can be distinguished [10]. Thus, the de-convolution of neutrons seems to be possible. If the method is accomplished, the time resolution for higher-energy neutrons (>10 keV) at the N-mode will follow that of the D-mode 1. The simulated time resolutions at different operation modes are shown in Fig. 3. They indicate that the time resolution is better than 1% for the whole energy range of 1 eV – 100 MeV with the D-mode 2 and it is comparable with n-TOF for neutron's energy higher than 100 keV.

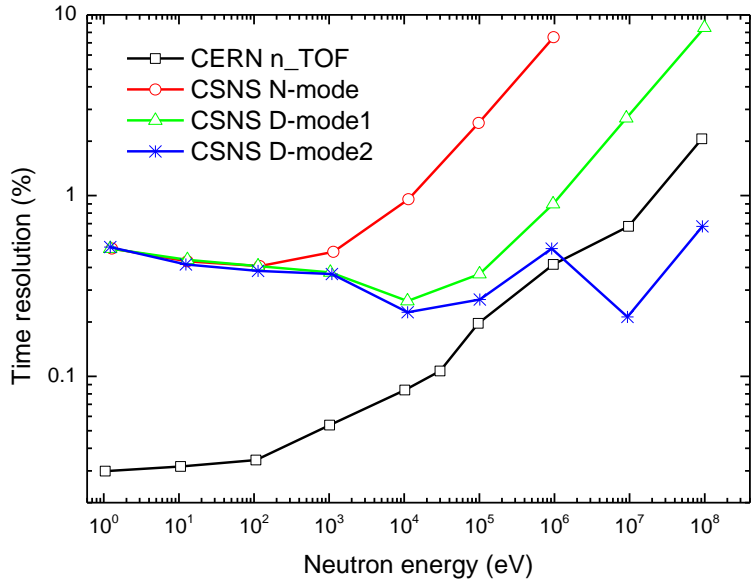

**Fig. 3 Time resolution comparison between CERN n-TOF and CSNS Back-n on different modes at 80 m**

## 3. Layout of the Back-n beam line and neutron collimation
### *3.1 Layout design*

The back-streaming neutrons emitting from the spallation target are transported in an 80-m long vacuum pipe along the reverse direction of incident proton beam. The first 20-m channel is common for both the proton and the neutron beam. At about 20 m from the target, a bending magnet which separates the proton and neutron beam can sweep off the charged particles contaminated in the neutron beam, and a shielding concrete wall at a distance of 24 m separates this high-radiation area from other parts of the tunnel. After the shielding wall, the first component in the Back-n beam line is a neutron beam window (NBW). Along the flight path, one neutron shutter (also functions as a collimator) and two collimators are used to control the neutron beam intensity and beam spot sizes, which are discussed in detail in the next section. At about 56 m and 76 m from the spallation target, two endstations (ES#1 & ES#2) will host seven detector systems (or spectrometers) in total. However, with only one set of them will be used in a time for experiments. ES#1 and ES#2 are used for high-flux and high-resolution experiments, respectively. A preparation room is allocated for experiments and temporal storage of detectors. Moreover, there is a dedicated Back-n control room in the RTBT building which is just above the ES#2. Due to limited space, an in-room and complex neutron dump is located at the rear side of ES#2. In order to suppress the background neutrons and gammas efficiently, several thick concrete walls and inner bushings with boron-containing polyethylene are set along the flight tunnel. Fig. 4 shows the layout of the Back-n beam line, and the distances of key components from the CSNS target.

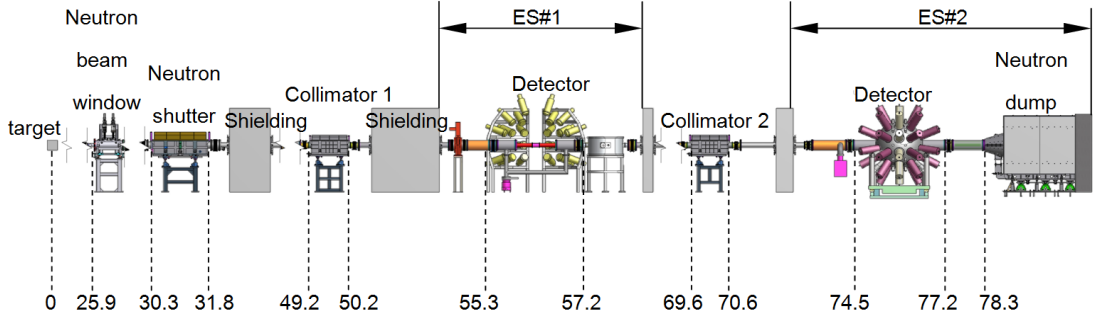

**Fig. 4. Layout of the Back-n beam line from the spallation target to the beam dump (distances are given in meters).**

### 3.2 Neutron collimation
### 3.2.1 Material and length of collimators

For simplicity，the transmission of neutron penetration through shields is represented by transmission efficiency [14-15]. The transmission efficiency (*T*) for neutrons in a single-element material is defined as

$$T = e^{-\sigma N_A \frac{\rho d}{A}}, \qquad (1)$$

where $N_A$ is the Avogadro constant, *A* is the atomic number of the material, *d* is the thickness in cm, *ρ* is the density in g/cm³, and *σ* is the reaction cross section in barn. The transmission efficiency (*T*) for neutrons in a compound material is defined as

$$T = \prod_{i=1}^{n} e^{-\sigma_i N_A \frac{\rho d k_i}{A_i}}, \qquad (2)$$

where $k_i$ is the mass percentage of each element component.

The multiple peaks and dips of the transmission spectrum in Fig. 5(a) arise from some characteristics for white neutrons scattering by atoms. Therefore, one simple way to stop the wide-energy neutrons is to use a combined material as collimators or shutter.

Fig. 5 shows the comparison of transmission efficiencies for several cascaded metal plates (Cu-Fe, Fe-Pb, Cu-Pb) with the Back-n beam. Taking the shielding effect and mechanical strength and cost into account, the combination of Cu-Fe is chosen as the material for the shutter and collimators. It can be seen from Fig. 5(c-d), that the transmission efficiencies of a combined block Cu-Fe with the thickness of 1.0 m and 1.5 m are sufficient to shield the white neutrons. Therefore, the lengths of the neutron shutter and two collimators are selected as 1.5 m and 1.0 m, respectively, which satisfy the requirements for the shielding and background control.

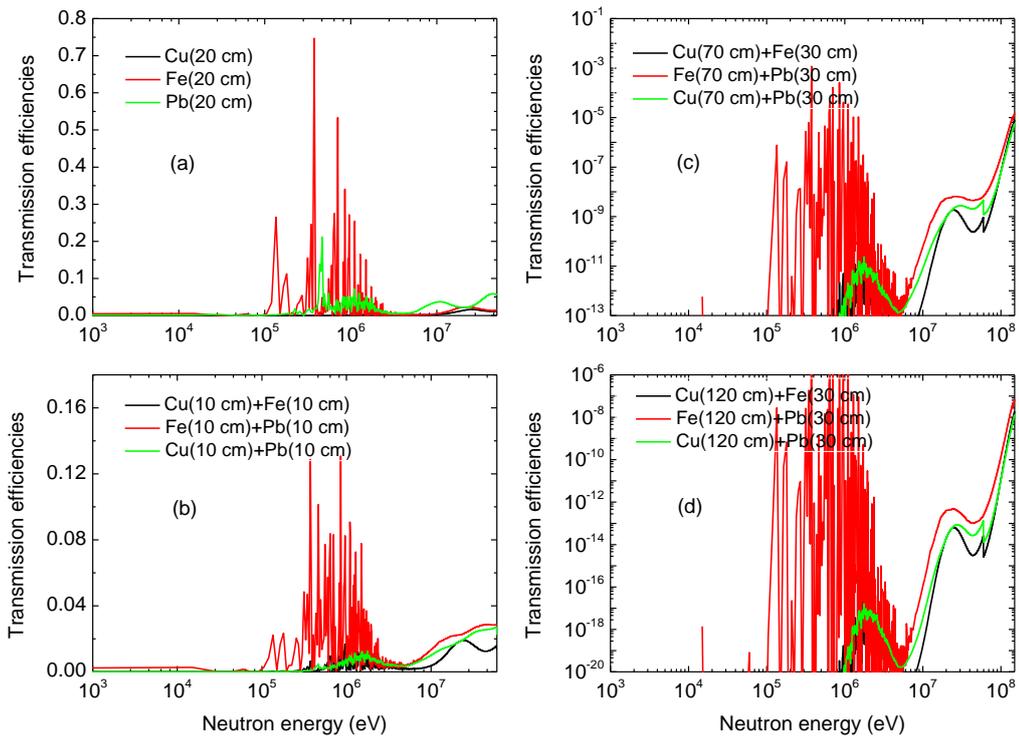

**Fig. 5 Comparison of the transmission efficiencies for different thickness and material combinations with a white neutron beam**

*3.2.2 Neutron optics*

According to the requirements of different experiments, three beam-spot settings, Φ30 mm, Φ60 mm and 90 mm × 90 mm, are designed at ES#2. In order to provide sharp profiles and reduce the background, a multi-stage collimation system has been used. The three collimators of different remote-controlled apertures, each followed by a shielding wall, are placed at the NBW before ES#1 and ES#2, as shown in Fig. 4. In order to save space and cost, the first collimator is also used as a shutter to stop neutrons from entering the endstations when a blank aperture is used.

A simplified optical schematic of neutrons is shown in Fig. 6. The neutron optical lines 1, 2 and 3 define the beam spot size, beam spot core and beam intensity, respectively. The shutter and Collimator 2 act as main collimators to control beam intensity and beam spot size. As an auxiliary collimator, Collimator 1 is used to reduce the beam halo and the background. The spot sizes at ES#1 are determined by the aperture setting of the shutter and Collimator 1 by following that of ES#2 in a passive manner.

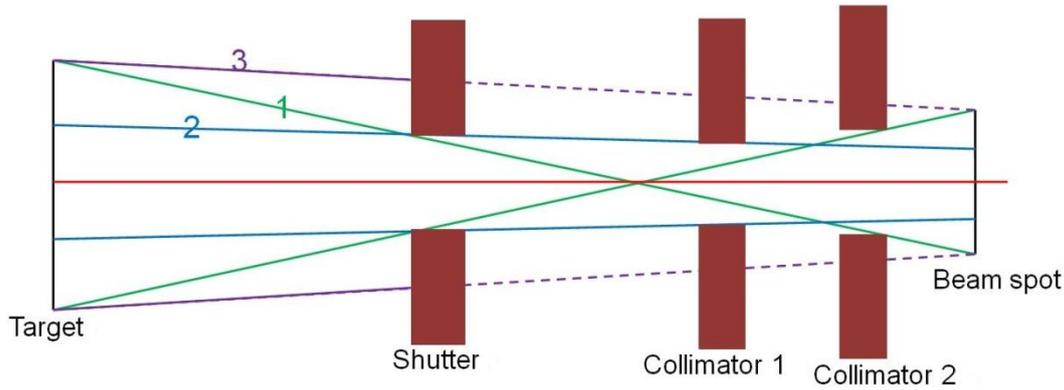

**Fig. 6 A simplified depiction of the neutron beam optics**

For the nominal beam spot of Φ30 mm at ES#2, the size of beam spot is smaller than the proton beam spot at the target. The shutter and Collimator 2 can easily define the beam spot according to line 1. In order to get a better beam uniformity, the shutter aperture can be increased slightly. On the other hand, optimization of the collimator apertures has an important influence on both beam intensity and uniformity. Collimator 1 is important to control the beam halo in practical cases. According to the optical relation, the aperture range of Collimator 1 can be defined by line 2 and 3. Thus the aperture setting for the three collimators can be determined. The nominal beam spot sizes of Φ60 mm and 90 mm × 90 mm are bigger in the vertical direction than the height of proton beam spot (40 mm) at target. Thus the neutron optics is slightly different as the target size also plays a role here. However, the beam halo and neutron background are also important for many experiments, and they also affect the optimization of the collimator apertures as discussed in the next sub-sections. Based on the optical results and the practical collimator designs, the three sets of collimation apertures together with the beam intensities are given in Table 1.

**Table 1 Back-n neutron beam spots & fluxes and corresponding collimation aperture parameters (@100 kW)**

| ES#2 spot (mm) | Shutter (mm) | Coll#1 (mm) | Coll#2 (mm) | ES#1 spot (mm) | ES#1 flux (/cm$^2$/s) | ES#2 flux (/cm$^2$/s) | Foreseen experiments |
|---|---|---|---|---|---|---|---|
| Φ30 | Φ12 | Φ15 | Φ40 | Φ20 | 2.20×10$^6$ | 7.81×10$^5$ | (n, t), (n,γ), (n, n'γ), (n,2nγ) |
| Φ60 | Φ50 | Φ50 | Φ58 | Φ50 | 4.33×10$^7$ | 1.36×10$^7$ | (n, f), (n, lcp), PFNS |
| 90×90 | 78×62 | 76×76 | 90×90 | 75×50 | 5.98×10$^7$ | 2.18×10$^7$ | neutron radiography |

*3.2.3 Simulations of collimation*

The most important components of collimation along the neutron flight path are the neutron shutter (also functions as a collimator) and the two collimators. Following the optical design, the optimization of the collimators has been carried out by multi-particle simulations. Due to very long flight path of 80 m, a secondary sampling of the neutron beam at the NBW has been used to save simulation time. In addition, to help find a first approximate solution, the simulations are started with black-body material for collimators and ducts. The black-body assumption is just limited to firstly study the relations between collimator aperture size and beam uniformity and intensity. And then many simulations and optimizations with more realistic conditions are followed.

In order to find balanced designs for beam intensity and uniformity for the given spot sizes, many aperture combinations have been used. For example, Fig. 7 shows the beam intensity dependence on the shutter aperture, and Fig. 8 shows the uniformity dependence on the shutter and collimator apertures. The uniformity is defined as the relative change of the beam intensity at the beam center ($I_{center}$) to that at the beam edge ($I_{edge}$) ($U = \frac{I_{center} - I_{edge}}{I_{center}} \times 100\%$). It can be seen that the uniformity has different trends with respect to apertures in the horizontal and vertical directions because of the asymmetry of the proton beam spot. Fig. 9 shows the neutron distributions for the three beam spots in the experimental halls, using the aperture settings of Table 1.

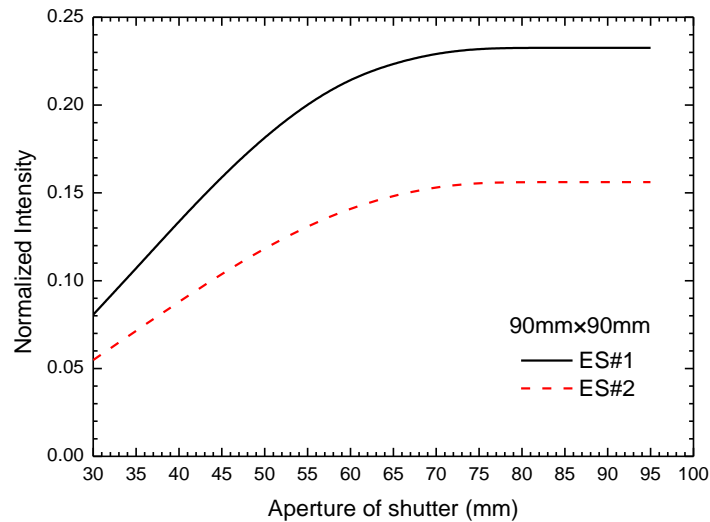

**Fig. 7 Neutron intensities of two endstations vs. shutter aperture by fixing the apertures of Collimator 1 & 2 (76 mm & 90 mm, respectively).**

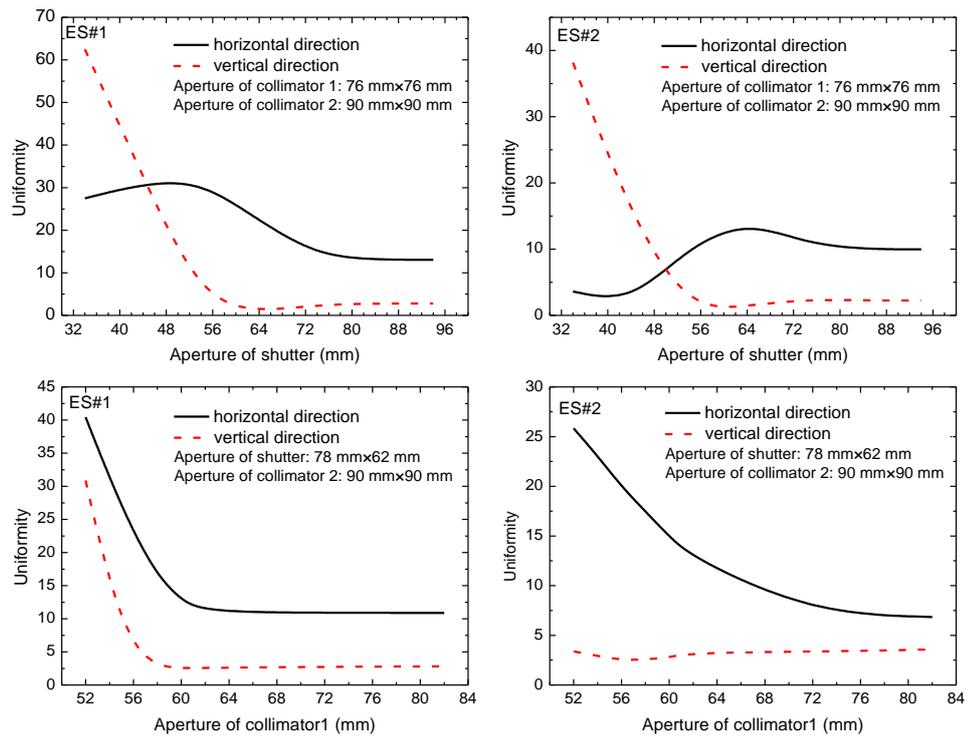

**Fig. 8.** Neutron beam-spot uniformity for 90 mm × 90 mm beam spot in the horizontal and vertical directions in ES#1 & ES#2 vs. apertures of shutter and collimator.

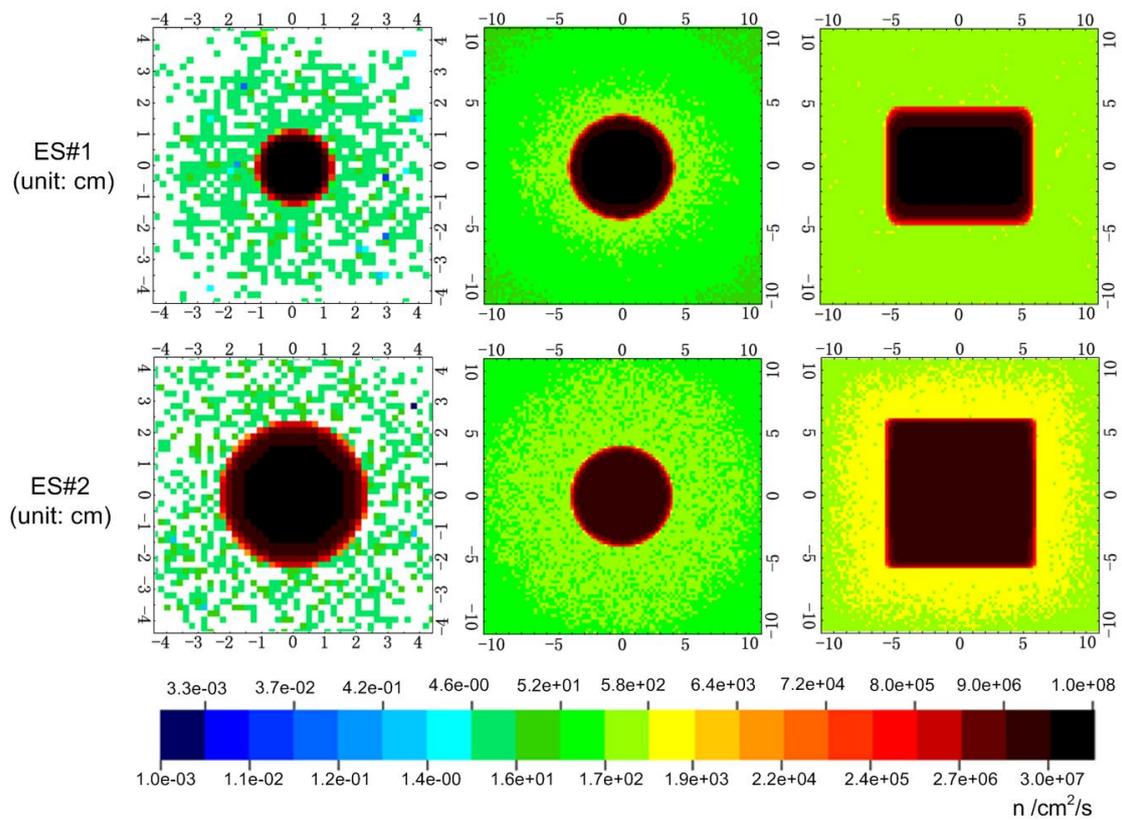

Fig. 9 Neutron density distribution for three cases at ES#1 & ES#2 (From left to right: Φ30 mm, Φ60 mm, 90 mm ×90 mm).

### 4. Background at the experimental halls

For nuclear data measurements at a white neutron source, it is an important issue to control the background at the experimental endstations. The background includes gammas, neutrons and charged particles. As mentioned before, the charged particles at the Back-n are well suppressed by the 15° bending magnet in the RTBT beam line. Therefore, only gamma and neutron backgrounds are a matter of concern here. They mainly arise from the proton beam loss in the nearby RTBT and spallation target, leakage of the collimators and dump, and also from spectrometers. The experiment related background varies with individual experimental setup and is not discussed in this paper. As the beam spot of 90 mm × 90 mm is mainly used for neutron imaging, the background is much less important for nuclear data measurements. Table 2 shows the contribution of different main source terms to the experimental hall backgrounds.

Table 2 Background in the experimental halls for the beam spot of Φ30 mm (fluence out/in beam, @100 kW)

| Background source | Background at ES#1 | | Background at ES#2 | |
|---|---|---|---|---|
| | Neutron | Gamma | Neutron | Gamma |
| RTBT(proton beam loss) | $\sim 10^{-7}$ | $\sim 10^{-7}$ | Negligible | Negligible |
| Target/collimators/pipe | $\sim 10^{-6}$ | $\sim 10^{-6}$ | $\sim 10^{-8}$ | $\sim 10^{-8}$ |
| Neutron dump | Negligible | Negligible | $\sim 10^{-9}$ | $\sim 10^{-9}$ |

*4.1 From proton beam loss*

The proton loss in the nearby RTBT tunnel will produce many neutrons and gammas. A part of high-energy neutrons will leak into the experimental hall through the shielding wall which separates the two tunnels. Although primary gammas can be stopped by the shielding wall, secondary gammas will be produced concomitantly by neutrons.

In order to suppress the neutrons penetrating the shielding wall, the first step is to insert an 80 cm thick iron block in the 2-m thick concrete shielding wall in the section separating ES#1. Thereafter, the usual requirement of 1 W/m for beam loss in the RTBT is reduced to 0.1 W/m in this special section, and this can be achieved as the RTBT section is a straight section composed of simple periodic focusing transport units. Moreover, a wall-cloth suit of 5-cm thick boron-containing polythene is installed on the four walls and ceil at ES#1. The neutron and gamma backgrounds in the experimental halls by the proton beam loss are simulated by FLUKA. A uniformly distributed line-type loss source of 0.1 W/m in RTBT is assumed and the neutron and gamma fluence distributions are shown in Fig. 10. With these actions, both the neutron and gamma backgrounds can be suppressed to an acceptable level at ES#1, whereas the background contribution at ES#2 from proton loss is totally negligible.

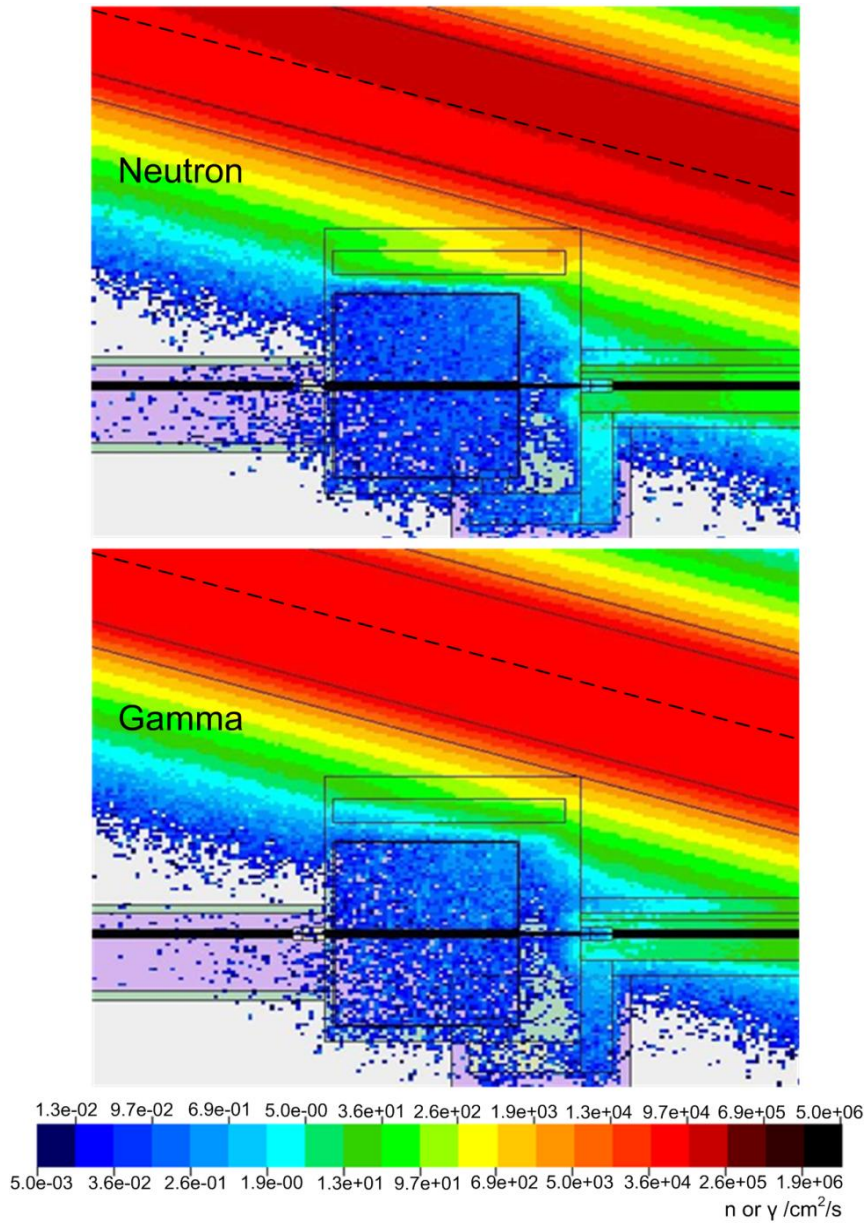

**Fig. 10** Neutron & gamma backgrounds in the ES#1 experimental hall. The dotted line represents the line-type proton loss source.

*4.2 From incoming neutrons*

On their way to the endstations, the back-streaming neutrons and gammas pass through the NBW, shutter, collimators and shielding walls along the neutron beam line, and produce scattered neutrons and gammas. Obviously, this part of the background contribution is complicated and can be classified as follows: (1) neutrons and gammas due to leakage from the collimators and front shield wall (2) scattered neutrons from the inner surface of the collimators and beam tubes (3) secondary gammas produced by neutrons interacting with shielding walls and (4) original gamma flash from the spallation target, which can be distinguished from the neutron burst by its early arrival at the endstations and its detection in the upstream beam line can be employed as the T0 signal for the TOF method. Fig. 11 shows the neutron and gamma backgrounds for three beam spots in the two endstations, simulated by FLUKA. In addition, with the help of inner bushings of boron-containing polyethylene in the shielding walls, the backgrounds in the experimental endstations can also be suppressed. For example, for the beam spot of Φ30 mm, the background is lower than the neutron beam flux by ~$10^7$–$10^8$ at ES#1 and ES#2.

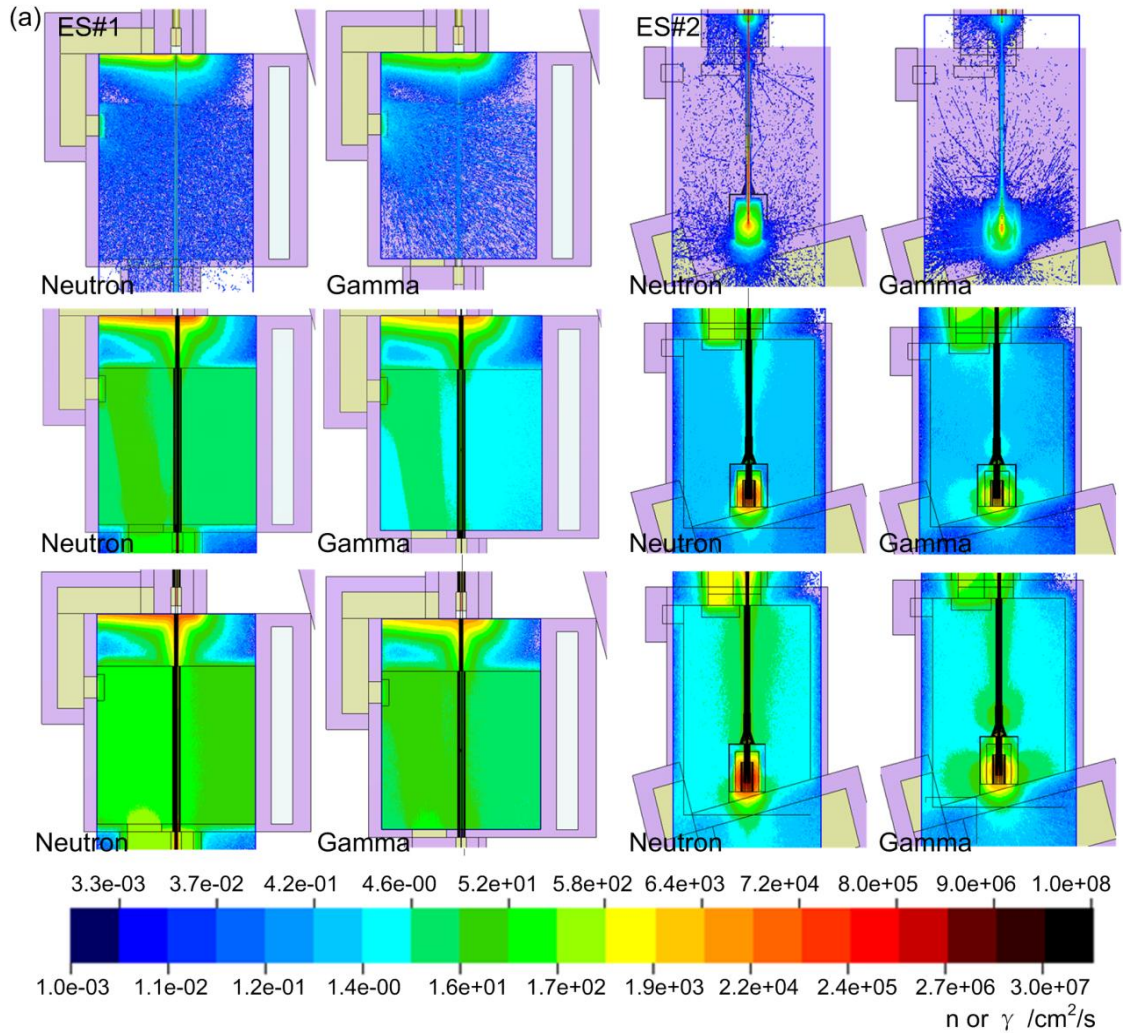

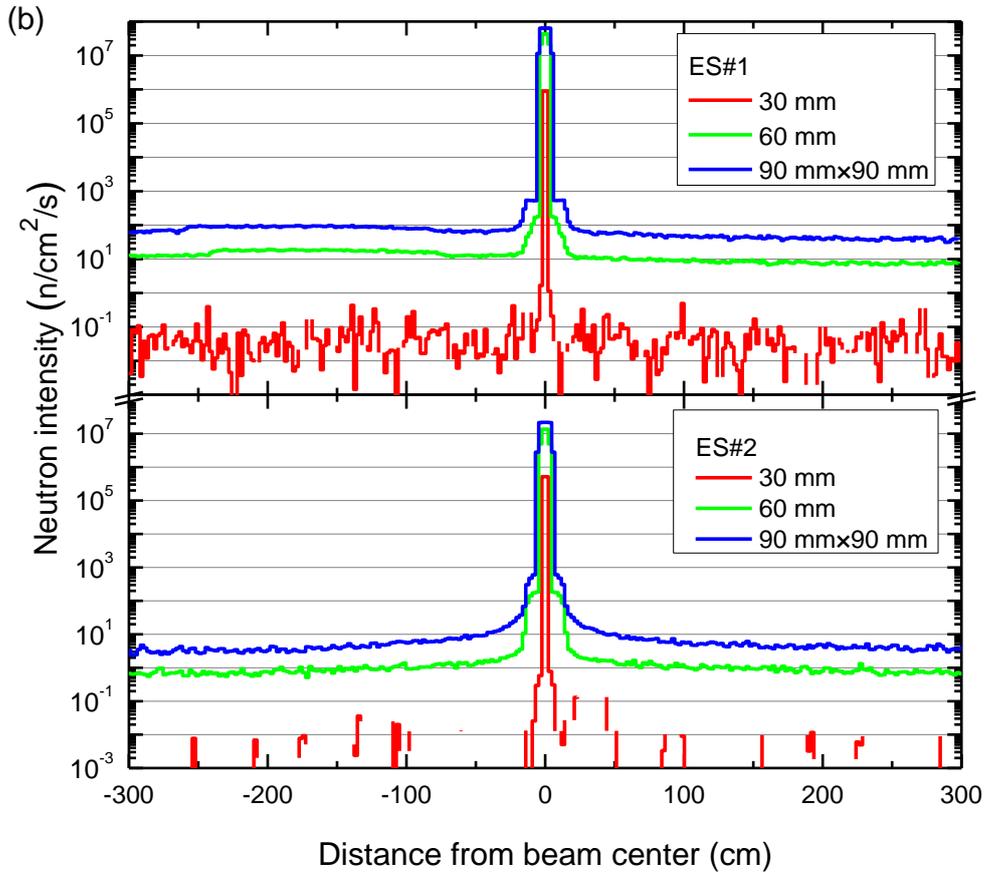

Fig. 11 Neutron beams and backgrounds at 100 kW: (a) Neutron & gamma fluences (From up to down: Φ30 mm, Φ60 mm, 90 mm × 90 mm); (b) and neutron profiles at ES#1 and ES#2

*4.3 From neutron dump*

Due to the space limitation of the CSNS layout, the neutron dump needs to be placed at the rear side of the ES#2. The backscattered and leaked neutrons and secondary gammas from the dump will contribute to the ES#2 backgrounds. In order to suppress this part of background, a sophisticated in-room compact dump has been designed [16]. For this study, the neutron beam is sampled again in FLUKA at the dump entrance. Fig. 12 shows the neutron background due to dump in the ES#2 experimental hall. The off-beam background is about $10^{-8}$, relative to the neutron flux for the small beam spot of Φ30 mm. The in-beam background is about three orders of magnitude higher than the off-beam background. The gamma background is almost equivalent to that of the neutron. For the other two cases, the backgrounds meet the experimental requirements.

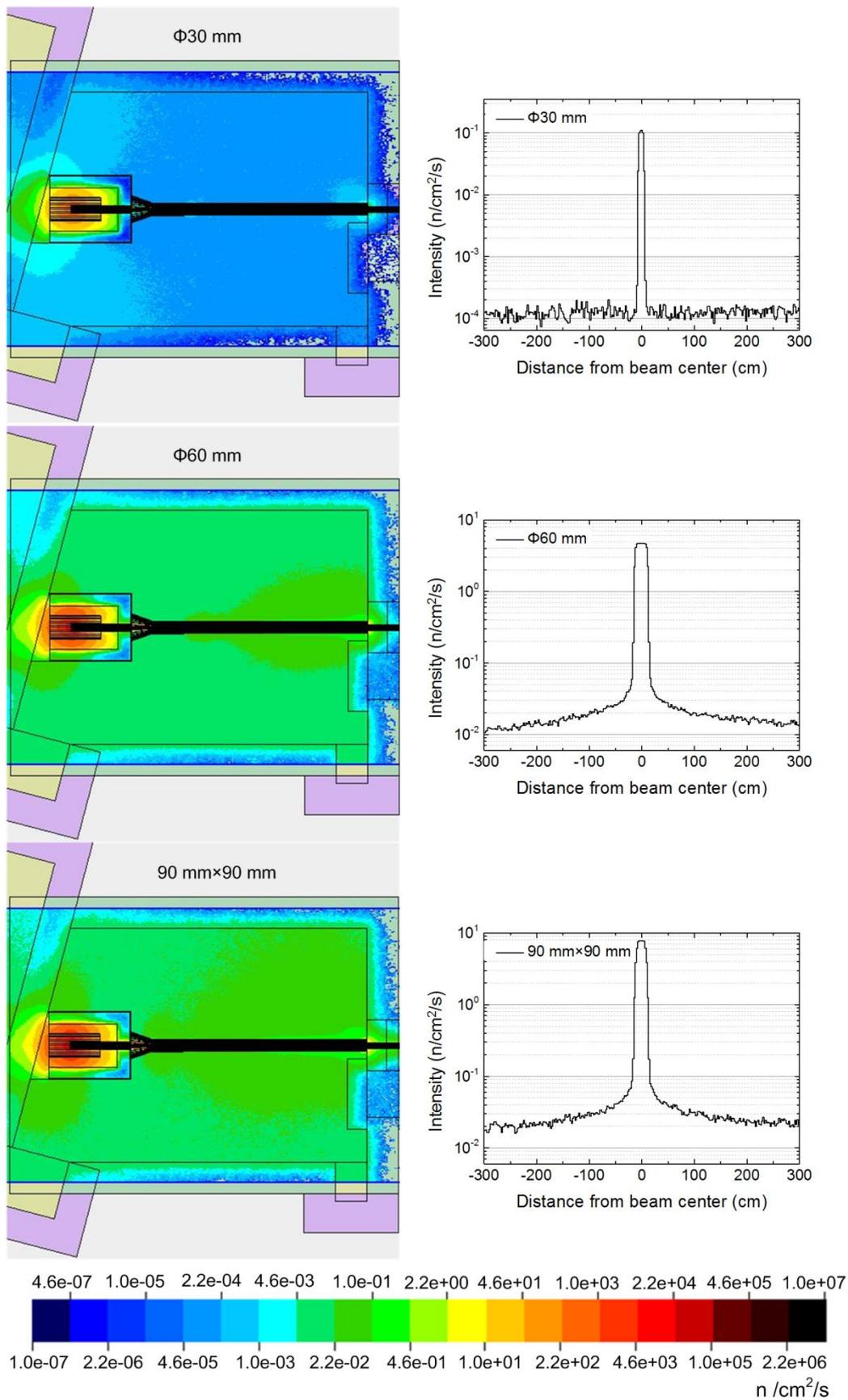

**Fig. 12 Neutron backgrounds from the dump contribution in the ES#2 hall for three neutron beam spots**

## 5. Key device designs
### *5.1 Neutron beam window*

NBW is located at a distance of 26 m from the target, which separates the vacuum conditions between the proton beam line ($10^{-6}$ Pa) and neutron beam line ($10^{-4}$ Pa) and also allows riskier vacuum down in the neutron beam line. For the window material, it is important that it produces as little distortion as possible to the energy spectrum shape and to beam intensity. In Fig. 13, several window materials are compared by using the energy-dependent transmission efficiencies in the whole neutron energy range. Finally, aluminum alloy A5083 of 1 mm thickness has been chosen for its good mechanical and neutron transmission properties. An additional advantage is that this material is also used for the proton beam window. A double spherical shape allows the thin window to stand for 1-atm pressure. At the same time, in the same vacuum chamber, a remote-controlled Cd filter and/or another filter can be inserted into the beam path as a low-energy (<1 eV) neutron absorber or for in-beam background measurement.

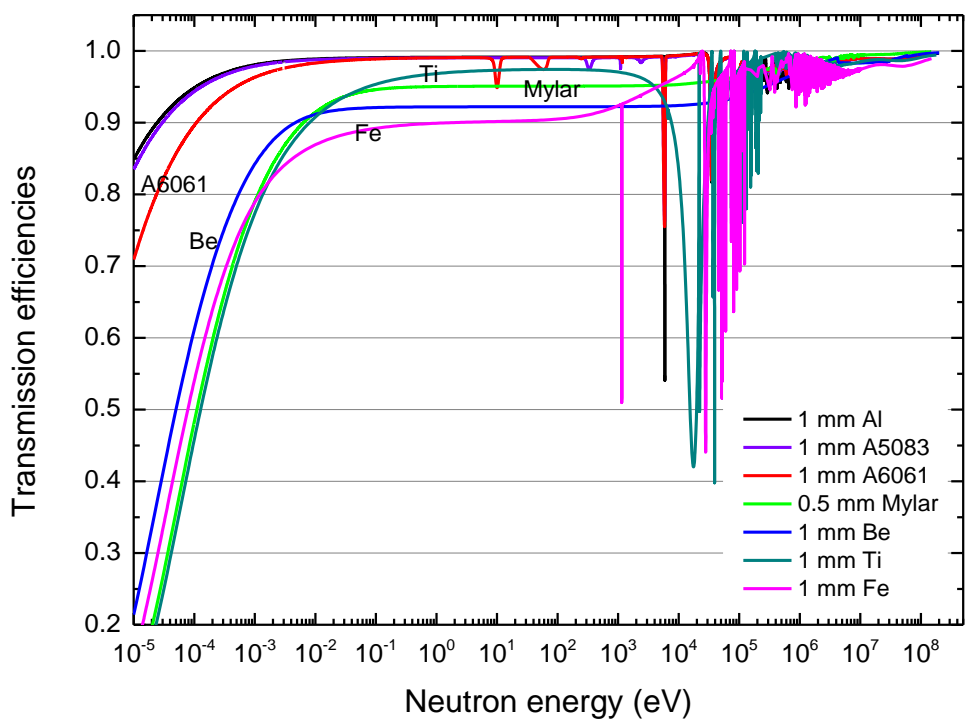

Fig. 13 Comparison of the transmission efficiencies for different materials in the whole neutron energy range

### *5.2 Neutron beam shutter*

The neutron shutter is located at a distance 31 m from the target. The shielding block is housed in a coffin-like vacuum chamber, and the block size is 1500 mm (length) × 150 mm (height) × 600 mm (width). As mentioned earlier, a combined material of 1.2-m copper and 0.3-m iron is used for better stopping of neutrons. Five positions in the horizontal direction are important for the remote-controlled shutter function: one blank position to stop neutrons, three apertures as the primary collimator for main experiments and a very small aperture of 3 mm for special low intensity applications. Magnetic fluid seal is used for mechanical motion in vacuum to shorten the lateral dimension. The shutter is also used as the key device in the radiation safety control.

*5.3 Collimators*

The two collimators are located before the two experimental endstations, followed by shielding walls. They have similar shutter structure, but with shorter length of 1 m or 0.7 m copper and 0.3 m iron. They require only three positions in the horizontal direction to design the collimation apertures. The two collimators are of slightly different dimensions: 1000 mm (length) × 150 mm (height) × 375 mm (width) for Collimator 1, 1000 mm (length) × 150 mm (height) × 385 mm (width) for Collimator 2.

*5.4 Neutron dump*

Due to general CSNS layout constraints, the neutron dump has to be placed inside the ES#2. In order to minimize its background contribution in the ES#2 hall, a sophisticated dump has been designed [16], as shown in Fig. 14. Copper and iron are used for stopping fast neutrons, boron polyethylene and lead shell absorb slow neutrons and secondary gammas respectively, and the cone-shape boron-polyethylene enclosing the incoming beam duct reduces the back-scattered neutrons. There are also three sets of changeable bushings in boron-polyethylene, corresponding to three beam spots, in the end section of the vacuum beam duct.

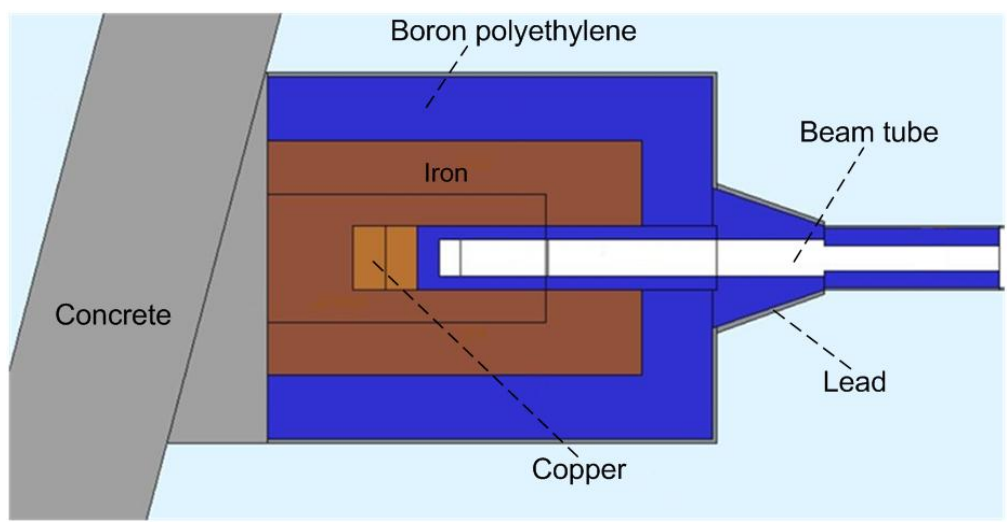

**Fig. 14 Geometry structure of the neutron dump.**

## 6. Conclusions

The back-streaming white neutron beam line at CSNS is being constructed for nuclear cross-section measurements. In addition to the simulation and analysis of the neutron energy spectrum and time resolution, the beam line design including optical design, neutron beam profile shaping and neutron/gamma background simulations is based on FLUKA Monte Carlo simulations. The simulations, which were carried out in the most realistic conditions achievable, show that the beam characteristics and backgrounds meet the design requirements of the Back-n layout and are very much suitable for nuclear data measurements. Some technical design aspects of the key elements are also discussed.

The main components of the beam line are under installation, and four detector systems are simultaneously under construction. The first batch of experiments or day-one experiments on nuclear data measurements are expected to be conducted from late 2017 to early 2018.

**Acknowledgements**

This work was supported by the National Key Research and Development Program of China (Project: 2016YFA0401600), National Natural Science Foundation of China (Project 11235012), the CSNS Engineering Project and Opening Special Foundation of State Key Laboratory of Intense Pulsed Radiation Simulation and Effect (Project number SKLIPR.1517). The authors would like to thank the colleagues of the CSNS white neutron source collaboration for discussions.